\documentclass[onecolumn,showpacs,preprintnumbers,amsmath,aps]{revtex4}
\usepackage{graphicx}
\usepackage{dcolumn}
\usepackage{bm}
\begin{document}
\def\eq#1{(\ref{#1})}
\def\fig#1{\ref{#1}}
\title{
---------------------------------------------------------------------------------------------------------------\\
Specific heat and thermodynamic critical field for the molecular metallic hydrogen}
\author{R. Szcz{\c{e}}{\`s}niak, M.W. Jarosik}
\affiliation{Institute of Physics, Cz{\c{e}}stochowa University of Technology, Al. Armii Krajowej 19, 42-200 Cz{\c{e}}stochowa, Poland}
\email{jarosikmw@wip.pcz.pl}
\date{\today} 
\begin{abstract}
In the framework of the Eliashberg formalism the free energy difference between the superconducting and normal state for the molecular metallic hydrogen
was calculated. The pressure values $p_{1}=347$ GPa and $p_{2}=428$ GPa were taken into consideration. It has been shown, that together with the increase of the pressure, grows the value of the specific heat jump at the critical temperature and the value of the thermodynamic critical field near zero Kelvin: $\left[\Delta C\left(T_{C}\right)\right]_{p2}/\left[\Delta C\left(T_{C}\right)\right]_{p1}\simeq 2.33$ and $\left[H_{C}\left(0\right)\right]_{p2}/\left[H_{C}\left(0\right)\right]_{p1}\simeq 1.74$. Next, it has been stated, that  the ratio $\Delta C\left(T_{C}\right)/C^{N}\left(T_{C}\right)$ also increases from $1.91$ to $2.39$; whereas $T_{C}C^{N}\left(T_{C}\right)/H^{2}_{C}\left(0\right)$ decreases from $0.152$ to $0.140$. The last results prove that the considered parameters significantly diverge from the prediction based on the BCS model.
\end{abstract}
\pacs{74.20.Fg, 74.25.Bt, 74.62.Fj}
\maketitle
%
\section{Introduction}

The study of the metallic hydrogen's properties has been lasting for over seventy years. In 1935, Wigner and Huntington for the first time suggested, that under the influence of high pressure ($p$) the hydrogen should transform into the molecular metallic phase [1]. The later theoretical results set the metallization of hydrogen in the pressures range from $300$ GPa to $400$ GPa \cite{Zhang}, \cite{Stadele}. It is worth mentioning, that the understanding of the high-pressure properties of hydrogen seems to be substantial due to the fact, that this element in the metallic state (both molecular and atomic) is appearing inside the planets of the Jovian type \cite{Fortney}.

The next step was made by Ashcroft who suggested, that the metallic hydrogen could be potential high-temperature superconductor \cite{Ashcroft}. Since that moment, the constant interest in the properties of the hydrogen's superconducting state has been dated. In particular, the numerical results predict that in the range of the "lower" pressures (up to $500$ GPa) the critical temperature ($T_{C}$) is of the order ($80$-$300$) K \cite{Zhang}, \cite{Richardson}, \cite{Caron}, \cite{Cudazzo}. For the extremely high pressure ($2000$ GPa) the superconducting state in the atomic metallic hydrogen has been studied in the papers \cite{Maksimov}, \cite{Szczesniak1}. It has been shown that the critical temperature decreases from $631$ K to $413$ K for $\mu_{C}^{*}\in(0.1,0.5)$, where $\mu_{C}^{*}$ denotes the critical value of the Coulomb pseudopotential. In the considered case the other thermodynamic parameters diverge from the BCS values \cite{BCS} e.g.: the dimensionless ratio $r_{1}\equiv \Delta C\left(T_{C}\right)/C^{N}\left(T_{C}\right)$ is changing from $1.82$ to $1.68$ together with the Coulomb pseudopotential's growth, whereas the minimum value of $r_{2}\equiv T_{C}C^{N}\left(T_{C}\right)/H^{2}_{C}\left(0\right)$ is equal to $0.162$ \cite{Szczesniak1}. The symbols defining the ratios $r_{1}$ and $r_{2}$ have following meaning: $\Delta C\left(T_{C}\right)$ denotes the specific heat difference between the superconducting and normal state at the critical temperature, $C^{N}\left(T_{C}\right)$ represents the specific heat of the normal state, while $H_{C}\left(0\right)$ is the value of the thermodynamic critical field at the temperature of zero Kelvin. 

In the literature the specific heat and the thermodynamic critical field were not determined for the molecular metallic hydrogen. Due to the large values of the electron-phonon constant ($\left[\lambda\right]_{p_{1}}=0.93$ and $\left[\lambda\right]_{p_{2}}=1.2$) it has to be presumed, that above quantities should be calculated in the framework of the Eliashberg formalism \cite{Eliashberg}. In the paper, we take into consideration the following values of the pressure: $p_{1}=347$ GPa and $p_{2}=428$ GPa. In this case the molecular metallic hydrogen crystallizes in the {\it Cmca} structure \cite{Zhang}, \cite{Cudazzo1}. 

\section{THE ELIASHBERG EQUATIONS}

The BCS theory is based on the Hamiltonian, which models the pairing interaction in the simplest effective way. We notice that the BCS Hamiltonian can be derived from the more realistic Fr\"{o}hlich's operator ($H_{F}$), which describes the electron-phonon coupling in the open form \cite{Frohlich}, \cite{TransKanon}. The Eliashberg equations are derived directly from $H_{F}$ with an use of the thermodynamic Green functions \cite{Elk}. As a result one can obtain \cite{Eliashberg}: 
\begin{widetext}
\begin{equation}
\label{r1}
Z_{n}=1+\frac{1}{\omega_{n}}\frac{\pi}{\beta}\sum_{m=-M}^{M}
                             \lambda\left(i\omega_{n}-i\omega_{m}\right)
                             \frac{\omega_{m}Z_{m}}
                             {\sqrt{\omega_m^2Z^{2}_{m}+\phi^{2}_{m}}}
\end{equation}
and
\begin{equation}
\label{r2}
\phi_{n}=
                                  \frac{\pi}{\beta}\sum_{m=-M}^{M}
                                  \left[\lambda\left(i\omega_{n}-i\omega_{m}\right)-\mu_{C}^{*}\theta\left(\omega_{c}-|\omega_{m}|\right)\right]
                                  \frac{\phi_{m}}
                                  {\sqrt{\omega_m^2Z^{2}_{m}+\phi^{2}_{m}}}.
\end{equation}
\end{widetext}
The solutions of the Eliashberg equations are two functions defined on the imaginary axis: the wave function renormalization factor ($Z_{n}\equiv Z\left(i\omega_{n}\right)$) and the order parameter function ($\phi_{n}\equiv\phi\left(i\omega_{n}\right)$); $\omega_{n}\equiv \left(\pi / \beta\right)\left(2n-1\right)$ is the $n$-th Matsubara frequency, where $\beta\equiv\left(k_{B}T\right)^{-1}$ ($k_{B}$ denotes the Boltzmann constant). In the framework of the Eliashberg formalism the order parameter is defined as: $\Delta_{n}\equiv \phi_{n}/Z_{n}$. The symbol $\lambda\left(z\right)$ represents the pairing kernel: 
\begin{equation}
\label{r3}
\lambda\left(z\right)\equiv 2\int_0^{\Omega_{\rm{max}}}d\Omega\frac{\Omega}{\Omega ^2-z^{2}}\alpha^{2}F\left(\Omega\right).
\end{equation}
The Eliashberg functions for the pressures $p_{1}$ and $p_{2}$ ($\alpha^{2}F\left(\Omega\right)$) were determined in the paper \cite{Zhang}. The symbol $\Omega_{\rm{max}}$ denotes the maximum phonon frequency, where $\left[\Omega_{\rm{max}}\right]_{p_{1}}=477$ meV and $\left[\Omega_{\rm{max}}\right]_{p2}=508$ meV.

The depairing correlations, appearing between electrons, are modeled with the help of the Coulomb pseudopotential $\mu_{C}^{*}$; the symbol $\theta$ denotes the Heaviside unit function and $\omega_{c}$ is the cut-off frequency ($\omega_{c}=3\Omega_{\rm{max}}$). In the paper we have assumed low value of the Coulomb pseudopotential for both considered pressures ($\mu^{*}_{C}=0.1$). The assumption above can be justified by referring to the Bennemann-Garland formula \cite{Bennemann}: $\mu_{C}^{*}\sim 0.26\rho\left(0\right)/\left[1+\rho\left(0\right)\right]$, where the symbol $\rho\left(0\right)$ indicates the value of the electronic density of states at the Fermi energy. In particular, we have: $\rho_{1}\left(0\right)=0.4512$ states/Ry/spin for $p_{1}$ and $\rho_{2}\left(0\right)=0.4885$ states/Ry/spin for $p_{2}$ \cite{Zhang}. Thus, $\left[\mu_{C}^{*}\right]_{p1}$ and $\left[\mu_{C}^{*}\right]_{p2}$ amounts $\sim 0.081$ and $\sim 0.085$ respectively.

From the mathematical point of view the Eliashberg set is composed of the strongly non-linear algebraic equations with the integral kernel $\lambda\left(z\right)$. In order to achieve stable solutions one needs to take into account adequately large number of the equations. In the paper we have assumed $M=800$, what assured stability of the solutions beginning from the temperature of $T_{0}=11.6$ K ($1$ meV). The Eliashberg equations were solved by using the iterative method presented in the papers \cite{Szczesniak2} and \cite{Szczesniak3}.

\section{THE NUMERICAL RESULTS}
%

\begin{figure}[h]%
\includegraphics*[scale=0.35]{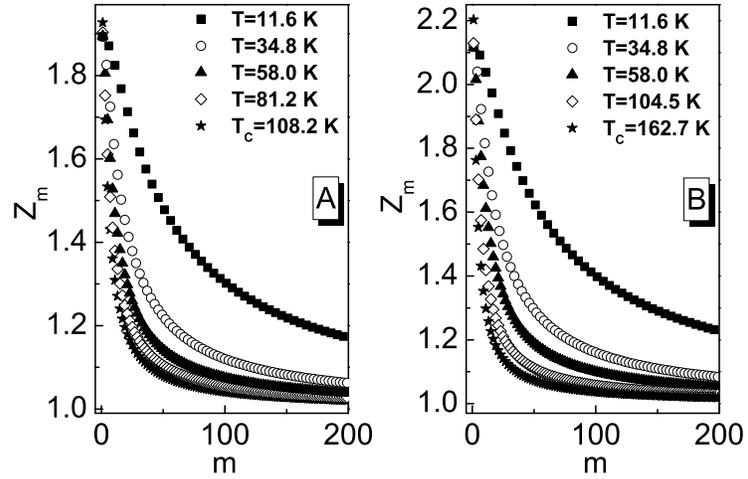}
\caption{The wave function renormalization factor on the imaginary axis for selected values of the temperature. The figure (A) shows results for $p_{1}$, the figure (B) for $p_{2}$.}
\label{f1}
\end{figure}

\begin{figure}[h]%
\includegraphics*[scale=0.35]{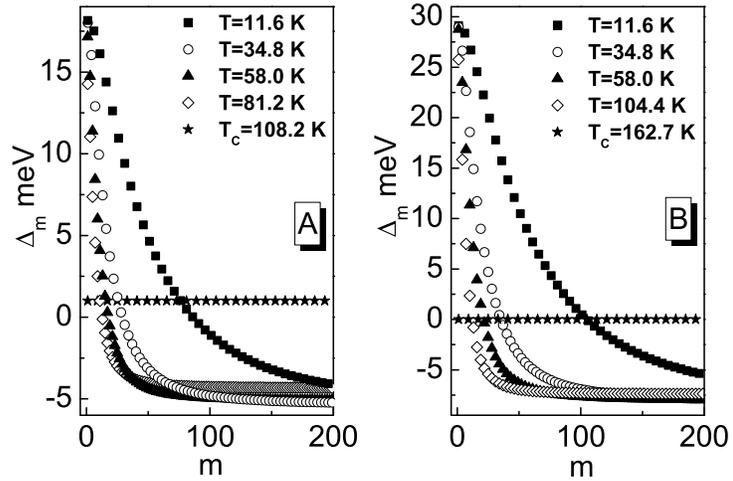}
\caption{The order parameter on the imaginary axis for selected values of the temperature. The figure (A) shows results for $p_{1}$, the figure (B) for $p_{2}$.}
\label{f2}
\end{figure}

The solutions of the Eliashberg equations for the selected temperatures have been presented in Figs. \fig{f1} and \fig{f2}. It can be easily noticed, that the functions $Z_{m}$ and $\Delta_{m}$ decrease together with the Matsubara frequencies' growth. However, $Z_{m}$ saturates considerably slower than $\Delta_{m}$.

The applied pressure significantly influences on the values of wave function renormalization factor and the order parameter. From the physical point of view the above fact means, that together with the increasing of $p$ increases the electron effective mass ($m^{*}_{e}\sim Z_{m=1}$) and the value of critical temperature ($\left[T_{C}\right]_{p_{1}}=108.2$ K,  $\left[T_{C}\right]_{p_{2}}=162.7$ K).

\begin{figure}[h]%
\includegraphics*[scale=0.35]{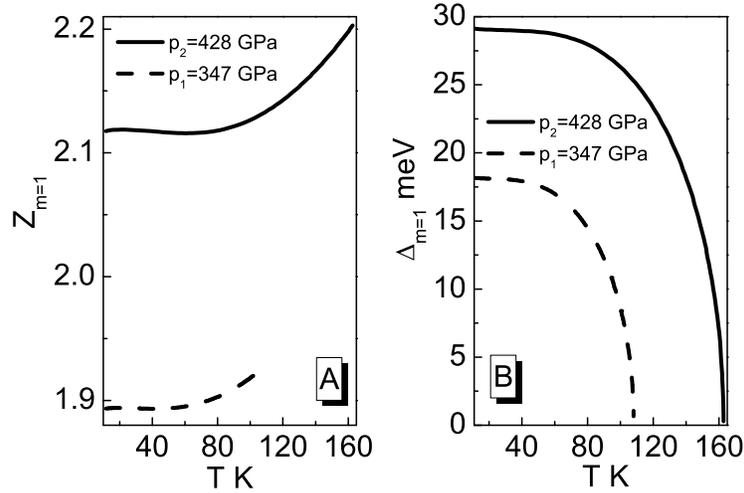}
\caption{
(A) The dependence of the wave function renormalization factor for the first Matsubara frequency on the temperature. (B) The dependence of the order parameter for the first Matsubara frequency on the temperature. In both cases the results for $p_{1}$ and $p_{2}$ are presented.}
\label{f3}
\end{figure}

Analyzing the dependence of $Z_{m}$ and $\Delta_{m}$ on temperature it has been stated, that the solutions of the Eliashberg equations very unlikely evolve with $T$. In Fig. \fig{f3} we have plotted the functions $Z_{m=1}\left(T\right)$ and $\Delta_{m=1}\left(T\right)$. The presented results show, that the wave function renormalization factor is weakly dependent on the temperature and takes its maximum for $T=T_{C}$. In contrast, the temperature dependence of the order parameter is strong and can be modeled by using the formula: $\Delta_{m=1}\left(T\right)=\Delta_{m=1}\left(T_{0}\right)\sqrt{1-\left(\frac{T}{T_{C}}\right)^{\beta}}$, where: $\left[\Delta_{m=1}\left(T_{0}\right)\right]_{p_{1}}=18.15$ meV, $\left[\Delta_{m=1}\left(T_{0}\right)\right]_{p_{2}}=29.12$ meV, $\left[\beta\right]_{p_{1}}=3.58$ and $\left[\beta\right]_{p_{2}}=3.61$. 

%
%
%
%

The thermodynamic properties of the molecular metallic hydrogen can be explicitly determined on the basis of the free energy difference between
the superconducting and normal state ($\Delta F$) \cite{Bardeen}:
\begin{equation}
\label{r4}
\frac{\Delta F}{\rho\left(0\right)}=-\frac{2\pi}{\beta}\sum_{m=1}^{M}
\left(\sqrt{\omega^{2}_{m}+\Delta^{2}_{m}}- \left|\omega_{m}\right|\right)
(Z^{{\rm S}}_{m}-Z^{N}_{m}\frac{\left|\omega_{m}\right|}
{\sqrt{\omega^{2}_{m}+\Delta^{2}_{m}}}),  
\end{equation}  
where the functions $Z^{S}_{m}$ and $Z^{N}_{m}$ denote the wave function renormalization factors for the superconducting (S) and normal (N) state respectively.

\begin{figure}[h]%
\includegraphics*[scale=0.35]{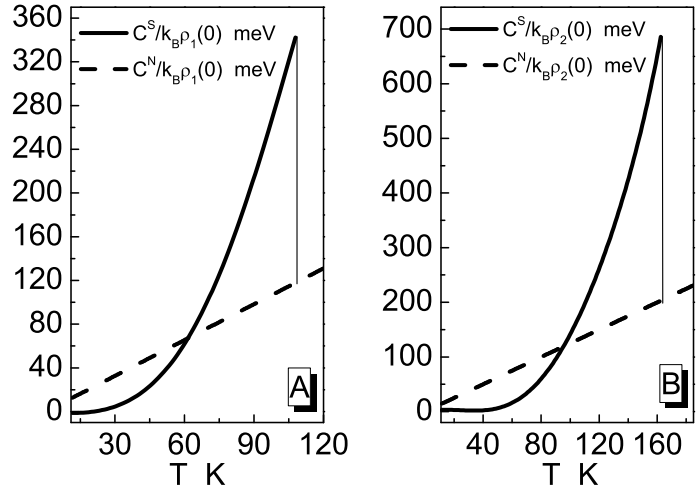}
\caption{
The dependence of the specific heat in the superconducting and normal state on the temperature. The figure (A) shows results for $p_{1}$, the figure (B) for $p_{2}$. The vertical line indicates a position of the specific heat jump that occurs at $T_{C}$.}
\label{f4}
\end{figure}

In the first step, on the basis of Eq. \eq{r4}, we have calculated the specific
heat difference between the superconducting and normal state $\left(\Delta C\equiv C^S-C^N\right)$: 
\begin{equation}
\label{r5}
\frac{\Delta C}{k_{B}\rho\left(0\right)}
=-\frac{1}{\beta}\frac{d^{2}\left[\Delta F/\rho\left(0\right)\right]}
{d\left(k_{B}T\right)^{2}}.
\end{equation}
Next, the specific heat in the normal state has been calculated with an use of the formula:
\begin{equation}
\label{r6}
\frac{C^{N}}{ k_{B}\rho\left(0\right)}=\frac{\gamma}{\beta}, 
\end{equation}
where $\gamma\equiv\frac{2}{3}\pi^{2}\left(1+\lambda\right)$. In Fig. \ref{f4} we have plotted the temperature dependence of the specific heat for the superconducting and normal state. Assuming previously given values of the electronic density of states it can be shown, that together with the growth of $p$ the specific heat's jump at the critical temperature very strongly increases. In particular, we have: 
$\left[\Delta C\left(T_{C}\right)\right]_{p_{2}}/\left[\Delta C\left(T_{C}\right)\right]_{p_{1}}\simeq 2.33$. 

%
%

\begin{figure}[h]%
\includegraphics*[scale=0.35]{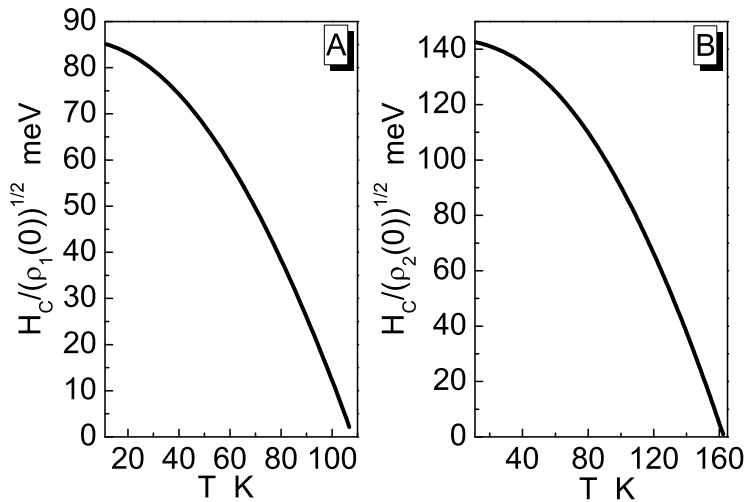}
\caption{
The thermodynamic critical field as a function of the temperature. The figure (A) shows results for $p_{1}$, the figure (B) for $p_{2}$.}
\label{f5}
\end{figure}

Below, we have calculated the values of the thermodynamic critical field (cgs units):  
\begin{equation}  
\label{r7}
\frac{H_{C}}{\sqrt{\rho\left(0\right)}}=\sqrt{-8\pi
\left[\Delta F/\rho\left(0\right)\right]}.
\end{equation}
In Fig. \ref{f5} we have presented the dependence of $H_{C}/\sqrt{\rho\left(0\right)}$ on the temperature. On the basis of obtained results we can see, that the value of the thermodynamic critical field near the temperature of zero Kelvin ($H_{C}\left(0\right)\simeq H_{C}\left(T_{0}\right)$) also strongly increases with the pressure: $\left[H_{C}\left(0\right)\right]_{p_{2}}/\left[H_{C}\left(0\right)\right]_{p_{1}}\simeq 1.74$.

%
%

On the basis of determined thermodynamic functions one can calculate two fundamental ratios: $r_{1}$ and $r_{2}$. Let us notice, that in the framework of BCS model these quantities have the universal values ($\left[r_{1}\right]_{\rm BCS}=1.43$ and $\left[r_{2}\right]_{\rm BCS}=0.168$) \cite{BCS}. 
For the molecular metallic hydrogen following results were obtained:
\begin{equation}
\label{r8}
\left[r_{1}\right]_{p_{1}}=1.91,\qquad \left[r_{1}\right]_{p_{2}}=2.39
\end{equation}
and
\begin{equation}
\label{r9}
\left[r_{2}\right]_{p_{1}}=0.152,\qquad \left[r_{2}\right]_{p_{2}}=0.140.
\end{equation}

It is easy to notice that the calculated ratios significantly diverge from the values predicted by the BCS theory. Additionally it should be underlined, that $r_{1}$ is increasing together with the pressure's growth, whereas the ratio $r_{2}$ is decreasing.

\section{SUMMARY}

In the paper the free energy difference between the superconducting and normal state for the molecular metallic hydrogen was calculated. The pressure values $p_{1}=347$ GPa and $p_{2}=428$ GPa were taken into consideration. On the basis of achieved results it has been shown, that the specific heat's jump at the critical temperature and the thermodynamic critical field near the temperature of zero Kelvin strongly increase together with the pressure's growth ($\left[\Delta C\left(T_{C}\right)\right]_{p2}/\left[\Delta C\left(T_{C}\right)\right]_{p1}\simeq 2.33$ and $\left[H_{C}\left(0\right)\right]_{p2}/\left[H_{C}\left(0\right)\right]_{p1}\simeq 1.74$). The obtained thermodynamic quantities enable the determination of the fundamental ratios: $r_{1}$ and $r_{2}$. It has been proven, that the ratios $r_{1}$ and $r_{2}$ very considerably differ from the values predicted by the BCS model. In particular, $r_{1}$ is increasing from $1.91$ to $2.39$ together with the pressure's growth; whereas $r_{2}$ is decreasing from $0.152$ to $0.140$.

\begin{acknowledgments}
The authors wish to thank Prof. K. Dzili{\'n}ski for providing excellent working conditions and the financial support. We also thank A.P. Durajski and D. Szcz{\c{e}}{\'s}niak for the productive scientific discussion that improved the quality of the presented paper. All numerical calculations were based on the Eliashberg function sent to us by: {\bf L. Zhang}, Y. Niu, Q. Li, T. Cui, Y. Wang, {\bf Y. Ma}, Z. He and G. Zou for whom we are also very thankful.
\end{acknowledgments}
%

%
\end{document}